\documentclass{ifacconf}
\usepackage{graphicx}      
\usepackage{natbib}        

\usepackage{xr}
\makeatletter
\newcommand*{\addFileDependency}[1]{
\typeout{(#1)}
\@addtofilelist{#1}
\IfFileExists{#1}{}{\typeout{No file #1.}}
}\makeatother

\usepackage{xcolor}

\usepackage{upgreek,bm}
\usepackage{amsfonts}
\newcommand{\THETA}{{\bm{\uptheta}}}
\newcommand{\X}{{\mathbf{x}}}
\newcommand{\Y}{{\mathbf{y}}}

\newcommand{\INDEX}{{t_i, y_i}}

\newcommand{\Yi}{y_{\INDEX}}

\newcommand{\YDATAi}{\bar{y}_{\INDEX}}

\newcommand{\THETAU}{{\THETA_{\mathbf{\mathrm{U}}}}}
\newcommand{\THETAM}{{\THETA_{\mathbf{\mathrm{M}}}}}

\newcommand{\UINPUT}{{\bm{\upnu}}}
\newcommand{\UINPUTX}{{\bm{\upnu}}(\X(t, \THETA))}
\newcommand{\UINPUTXU}{{\bm{\upnu}}(\X(t, \THETAU))}
\newcommand{\ULAYER}{{\bm{\upphi}}}
\newcommand{\U}{{\mathbf{U}}}
\newcommand{\N}{{\mathbf{N}}}

\newcommand{\XODE}{{\X_{\mathrm{ODE}}}}

\newcommand{\XNUDE}{{\X_{\mathrm{nUDE}}}}

\newcommand{\UINPUTXNUDE}{{\bm{\upnu}}(\XNUDE(t, \THETA))}

\usepackage{amsmath}

\newtheorem{theorem}{Theorem}
\newenvironment{proof}{\paragraph*{Proof:}}{\hfill$\square$}

\usepackage{mathtools}

\usepackage[
  pos={0.5\paperwidth,0.08\paperheight},
  firstpageonly=true,
  vanchor=t,angle=0,scale=0.5,color={[gray]{0.4}},
  text={This work has been submitted to IFAC for possible publication. Initial submission was March 18, 2024.}
]{draftwatermark}

\begin{document}

\begin{frontmatter}

\title{Non-Negative Universal Differential Equations With Applications in Systems Biology\thanksref{footnoteinfo}} 

\thanks[footnoteinfo]{
This work was supported by the Deutsche Forschungsgemeinschaft (DFG, German Research Foundation) under Germany’s Excellence Strategy (EXC 2047—390685813, EXC 2151—390873048) and under the project ID 432325352 – SFB 1454, by the German Federal Ministry of Education and Research (BMBF) under the CompLS program (GENImmune, grant no 031L0292F), and by the University of Bonn (via the Schlegel Professorship of J.H.).\\
Supplemental Material is available at https://doi.org/10.5281/zenodo.10834999. \\
\^{}\textit{Joint senior authors.\\
$^1$maren.philipps@uni-bonn.de\\
$^2$dilan.pathirana@uni-bonn.de\\
$^3$jan.hasenauer@uni-bonn.de}
}

\author[1]{Maren Philipps$^1$} 
\author[1]{Antonia Körner}
\author[1]{Jakob Vanhoefer} 
\author[1]{Dilan Pathirana\^{}$^2$} 
\author[1,2]{Jan Hasenauer\^{}$^3$}

\address[1]{Faculty of Mathematics and Natural Sciences, and the Life and Medical Sciences Institute (LIMES), Rheinische Friedrich-Wilhelms-Universit\"at Bonn, Bonn, Germany}
\address[2]{Computational Health Center, Helmholtz Zentrum München Deutsches Forschungszentrum für Gesundheit und Umwelt (GmbH), Neuherberg, Germany}

\begin{abstract}                
    Universal differential equations (UDEs) leverage the respective advantages of mechanistic models and artificial neural networks and combine them into one dynamic model. However, these hybrid models can suffer from unrealistic solutions, such as negative values for biochemical quantities. We present non-negative UDE (nUDEs), a constrained UDE variant that guarantees non-negative values. Furthermore, we explore regularisation techniques to improve generalisation and interpretability of UDEs.
\end{abstract}


\end{frontmatter}

\section{Introduction}

Systems biology aims to find a mechanistic understanding of biological processes and mathematical models are an important tool to achieve this. Mechanistic models are constructed to be consistent with the available information; however, their construction is limited by the current mechanistic understanding of the biological process. In practice, incomplete information is akin to multiple hypotheses about a process, which can be addressed by model selection. However, model selection can be time-consuming, particularly when there are large knowledge gaps.

A novel approach to address incomplete information is universal differential equations (UDEs), which combine dynamical mechanistic and machine learning (ML) models. UDEs represent known mechanisms explicitly, and unknown mechanisms by universal approximators like artificial neural networks (ANNs).
These hybrid models have been shown to require less training data and improve interpretability over purely data-driven ML \citep{karniadakis2021physics}.
An open issue is that UDEs can produce negative values, even for strictly non-negative quantities such as molecular concentrations or population sizes.
Dynamical mechanistic models, such as ordinary differential equations (ODEs), do not have this issue because the mechanisms can be chosen to ensure non-negativity.

Here, we present an extension to the UDE framework that ensures non-negativity (nUDE).
We provide a proof of the non-negativity, and evaluate nUDEs on a synthetic and a real-world example. Moreover, we introduce a new regularisation method to control the over-fitting of the ANN in (n)UDEs. We find that our non-negativity approach may bias the ANN training; however, this bias can be reduced, and calibration efficiency and model quality can be preserved, by choosing the non-negativity factor carefully.


\section{Modelling}

\subsection{Mechanistic modelling with ODEs}
Mechanistic modelling is facilitated by the conversion of domain knowledge into actionable mathematical expressions, which can be used to understand the modelled behaviour in a virtual setting.
A significant benefit over non-mechanistic modelling is the ability to predict behaviour that is not represented in the available training data.
However, model construction can be time-consuming.
As the exemplary models in this work are taken from the literature, this process is not further discussed here. Reviews of this process are available in the literature, e.g. \cite{villaverdeProtocolDynamicModel2022}.

In systems biology, ODEs are commonly used to describe the time-dependent rate-of-change of biological entities, such as proteins on the level of cells, or groups of individuals on the level of populations. Here, we consider a general form of ODEs with initial conditions, i.e., initial value problems (IVPs):
\begin{align}
\begin{rcases}
    \displaystyle\frac{d\X(t, \THETAM)}{dt} &= \mathbf{f}(\X(t, \THETAM), t, \THETAM), \\
    \X(t_0, \THETAM) &= \X_{\mathbf{0}}(\THETAM).
\end{rcases}
\label{equation:ode_system}
\end{align}
The system changes with time according to the vector field $\mathbf{f}: \mathbb{R}^{n_\X}\times \mathbb{R}^{n_\THETAM} \rightarrow \mathbb{R}^{n_\X}$. The initial position, at $t=t_0$, is the initial condition $\X_{\mathbf{0}}: \mathbb{R}^{n_\THETAM} \rightarrow \mathbb{R}^{n_\X}$. The model is parameterised by $\THETAM$, which can contain values such as population growth rate constants.

\subsection{Machine learning with ANNs}
In supervised machine learning, input-output pairs are used to train an unknown function that represents the input-output mapping. Some ANNs, such as multi-layer feedforward networks, are universal approximators in the limit case, meaning they are capable of approximating functions to arbitrary precision (see e.g. \cite{hornikMultilayerFeedforwardNetworks1989}).
These ANNs are structured into layers of \emph{neurons}, where each neuron $\psi$ of each layer $\ULAYER$ is the composition of one affine and (usually) one non-linear transformation function.
The fully-connected ANN $\U$ can then be understood as a composition of layers, i.e. 
\begin{equation*}
\U(\UINPUT, \THETAU)=\left(\ULAYER_L\circ\ULAYER_{L-1}\circ ... \circ \ULAYER_1\right)(\UINPUT, \THETAU),
\end{equation*}
where $\THETAU$ are the weight and bias parameters of the affine functions, $\UINPUT \in \mathbb{R}^{n_\UINPUT}$ is the input, and $L$ is the total number of layers \citep{goukRegularisationNeuralNetworks2021}.

Many commonly used activation functions are Lipschitz continuous \citep{goukRegularisationNeuralNetworks2021}. These include the hyperbolic tangent $\tanh(z)$, the logistic sigmoid $\left(1+\exp(-z)\right)^{-1}$, and the rectified linear unit (ReLU) $\max(0, z)$. 
As ANNs with these activation functions are thereby compositions of Lipschitz-continuous functions, these ANNs are also Lipschitz continuous; a property that we use to prove Theorem \ref{theorem:nudes}.

We describe the architecture of these ANNs by the number of neurons in each layer. For example, 3/3/2 is a fully-connected feedforward ANN with 3 neurons in the first and second layers, and 2 neurons in the output layer. Each neuron $i$ in layer $l$ outputs $\psi_{l,i} = \sum_j \mathcal{A}_{l,i,j}(w_{l,i,j} \psi_{l-1,j} + b_{l,i,j})$, where $j$ is the neuron index in the previous layer, $\psi_{l-1,j}$ is the output from neuron $j$ in the previous layer $j-1$, and $\mathcal{A}_{l,i,j}$, $w_{l,i,j}$ and $l_{k,i,j}$ are the activation function, weight and bias, respectively.
The output from layer $l$ with $n$ neurons is then $\ULAYER_l = \left(\psi_{l,1}, \cdots, \psi_{l,n}\right).$

\subsection{Universal differential equations}
Neural ODEs are ODEs similar to \eqref{equation:ode_system}, but with an ANN as their right-hand-side, i.e.
\begin{align*}
    \frac{d\X(t, \THETAU)}{dt} &= \U(\UINPUTXU, \THETAU),\\
    \X(t_0, \THETAU) &= \X_{\mathbf{0}}(\THETAU),
\end{align*}
where the input $\UINPUTXU$ is now some function of the state $\X$.

While mechanistic modelling and neural ODEs with ANNs both have important use cases, both approaches have drawbacks. UDEs have been introduced to exploit the strengths of each approach, and to enable modelling of partially unknown biological processes \citep{oliveira2004combining}. A general form for UDEs in different modelling formalisms is given in \cite{rackauckas2020universal}. In the ODE context, a formulation for UDEs is given by
\begin{align}
\begin{rcases}
    \displaystyle\frac{d\X(t, \THETA)}{dt} &= \mathbf{f}(\X(t, \THETA), t, \THETAM) + \U(\UINPUTX, \THETAU),\\
    \X(t_0, \THETA) &= \X_{\mathbf{0}}(\THETA),
\end{rcases}
\label{equation:ude_system}
\end{align}
where $\U$ enables modelling unknown process mechanisms in addition to the known mechanisms $\mathbf{f}$.
Here, $\THETA = (\THETAM, \THETAU)$, with mechanistic parameters $\THETAM$. Although $\U$ can be any universal approximator, in this study we only consider fully-connected feedforward ANNs.

We note that the system in \eqref{equation:ude_system} is not a universal approximator for a dynamical system, despite $\U$ being a universal approximator. However, this can be achieved by adding state variables to the system that are equipped with dynamics that are wholly-modelled in terms of a universal approximator \citep{dupont2019augmented}.

\subsection{Maximum likelihood estimation}
\label{methods:mle}
Mechanistic models, ANNs, and UDEs often contain unknown parameters, which can be estimated from data. One approach is to find the maximum likelihood estimate (MLE), which is the choice of parameter values $\THETA_{MLE}$ that maximises the probability of observing measurements $\mathbf{\bar y}$, i.e., the likelihood of the data under some system parameterised by $\THETA$. This requires an observation model $\mathbf{h}: \mathbb{R}^{n_\X} \rightarrow \mathbb{R}^{n_\Y}$ that maps model state space to data observation space. In general, measurements in the biological sciences are significantly noisy, hence the observables $\Y = \mathbf{h}(\X(t,\THETA),\THETA)$ are related to the data by $\bar{y}_\INDEX = y_\INDEX + \epsilon_\INDEX$, where $t_i \in {1,...,n_t}$ and $y_i \in {1,...,n_\Y}$ are used to index over measurements by timepoint and observable, respectively, and $\epsilon_\INDEX$ is measurement-specific noise.

For numerical efficiency, we minimise the negative log-likelihood function, 
$$J(\THETA)=\frac{1}{2}\sum_\INDEX \log(2\pi\sigma^2_\INDEX) + \frac{(\YDATAi - \Yi(\THETA))^2}{\sigma^2_\INDEX},$$
for i.i.d. Gaussian noise, i.e. $\epsilon_\INDEX \sim N(0, \sigma^2_\INDEX)$.

\section{Towards biologically meaningful UDEs}
A common issue in machine learning is over-fitting, characterised by the model adapting to noise or artefacts in the training data, which is deleterious for generalisation beyond the training domain \citep{ying2019overview}. 
This issue is more prominent in ML compared to mechanistic models that are constrained by domain knowledge. 
A special case of over-fitting in UDEs is the absorption of the dynamics that are encoded in the mechanistic terms ($\mathbf{f}$ in system \eqref{equation:ude_system}).
A broad spectrum of approaches to mitigate over-fitting have been introduced in ML, including early stopping, noise injection, and stochastic gradient descent. Few of these approaches have been transferred to the field of UDEs.
One common approach that we tested here is to regularise the system during training by introducing a \emph{learning bias} \citep{karniadakis2021physics}.

Another limitation of UDEs is that their dynamics are not necessarily biologically meaningful.
For example, a na\"ive function approximator would not adhere to the principles of mass conservation or non-negativity of biological quantities.
Just as mechanistic models can be designed to implicitly comply with such fundamental principles, mathematical constraints for ANNs can be used as an \emph{inductive bias} to strictly enforce biologically-reasonable model behaviour \citep{karniadakis2021physics}.

We describe the learning bias \emph{parameter regularisation}, introduce the learning bias \emph{output regularisation}, and introduce the inductive bias \emph{non-negative UDEs} (nUDEs).

\subsection{Parameter regularisation:}  
Parameter regularisation aims to reduce the magnitude of the parameters as a proxy for model flexibility. 
In particular, $\ell_1$ and $\ell_2$ norms are frequently used to directly penalise model parameters \citep{Goodfellow-et-al-2016}. 
Here, we use the $\ell_2$ (also known as Euclidean) norm of $\THETAU$, 
$\lVert \THETAU \rVert_2 = \sqrt{ \sum_i {\theta_{U_i}}^2 }$,
yielding the regularised objective 
\begin{equation*}
    J(\THETA) + \lambda_p \lVert \THETAU \rVert^2_2,
\end{equation*}
with regularisation parameter $\lambda_p \geq 0$.

\subsection{Output regularisation:} As parameter regularisation only indirectly limits the impact of $\U$ on the solution, we also consider a novel regularisation scheme, which we will refer to as output regularisation.
We compute non-zero contributions of $\U$ to the solution directly with
\begin{equation*}
R(\THETA) = \int_{t_0}^{t_f} \lVert \U_R(\THETA) \rVert_2 \,\mathrm{d}t,
\end{equation*}
where $\U_R=\U$ and $\U_R=\N\odot\U$ in the UDE and nUDE (see Section \ref{section:nudes}) cases, respectively. We set $t_f$ to the time of the last measurement in the training data 
and add the penalty into the regularised objective function as
\begin{equation*}
J(\THETA) + \lambda_o R(\THETA)^2,
\end{equation*}
with regularisation parameter $\lambda_o \geq 0$.

\subsection{Non-negative UDEs}
\label{section:nudes}
Regularisation in the objective function does not directly ensure desirable properties such as non-negativity for biological quantities. 
In this section, we present a formulation of a constrained UDE that ensures that state variables cannot become negative. We consider the model structure
\begin{align}
\small
    \frac{d\XNUDE(t, \THETA)}{dt} &= \mathbf{f}(\XNUDE(t, \THETA), t, \THETAM) \notag\\
    &\quad + \N(\XNUDE(t, \THETA)) \odot \U(\UINPUTXNUDE, \THETAU),
\label{equation:nude_system}
\end{align}
with Lipschitz-continuous functions $\U:\mathbb{R}^{n_\UINPUT}\rightarrow\mathbb{R}^{n_\X}$ and $\N:\mathbb{R}^{n_\X}\rightarrow\mathbb{R}^{n_\X}$, and choosing $\N:\lim_{x_i \rightarrow 0} N_i(\X) = 0\, \forall i \in \{1,\ldots,n_\X\}$ (e.g., $\N(\X)=\X$). $\odot$ is the element-wise (Hadamard) product. In the following, we present the properties of this non-negative UDE (nUDE). Some function inputs are omitted for brevity after their first use.

\begin{theorem}
Consider the ODEs IVP, 
\begin{align*}
\frac{d\XODE(t, \THETAM)}{dt} &= \mathbf{f}(\XODE(t, \THETAM), t, \THETAM),\\
\XODE(t_0, \THETAM) &\ge \mathbf{0},
\end{align*}
where $\mathbf{f}$ is Lipschitz-continuous. If the non-negative quadrant is invariant under $\mathbf{f}$, i.e. $\left.f_i\right|_{x_i=0} \ge 0\, \forall i \in \{1,\ldots,n_\X\}$, then $x_{nUDE,i}\ge 0\, \forall t \ge t_0$ in a nUDE system \eqref{equation:nude_system} with the same $\mathbf{f}$.
\label{theorem:nudes}
\end{theorem}

\begin{proof}
As the right-hand-side of \eqref{equation:nude_system} is composed of Lipschitz-continuous functions, the nUDE IVP has a unique solution $\XNUDE(t, \THETA)$, by the Picard–Lindel\"of theorem. The initial value $\XNUDE(t_0, \THETA)$ is non-negative. We will show that the solution remains non-negative, with a proof by contradiction.

Assume there exists some $i\in{1,...,n_\X}$ and $\tilde{t} > t_0$ s.t. $x_{\mathrm{nUDE},i}(\tilde{t}, \THETA) < 0$. As the initial condition is non-negative, there must be some $t^*\in[t_0, \tilde{t}]$ s.t. $x_{\mathrm{nUDE},i}(t^*,\THETA)=0$ and its derivative
\begin{equation}
    \left(\left.f_i + N_iU_i\right)\right|_{t=t^*} < 0.
    \label{proof:rhs_condition}
\end{equation}
As $\U$ and $\N$ are continuous $U_{\max}:=\max_{t\in [t_0, t^*]} U_i < \infty$ and $\left.N_i\right|_{x_{\mathrm{nUDE},i} = 0}= 0$. Hence, $|\N\odot\U|_i=|N_i\cdot U_i| \leq |N_i\cdot U_{\max}| = 0$ at $t = t^*$, and \eqref{proof:rhs_condition} simplifies to $\left.f_i\right|_{t=t^*} < 0$.
However, given $\left.f_i\right|_{x_i=0} \geq 0$, we arrive at the contradiction $\left.f_i\right|_{t=t^*} \geq 0$, hence no such $\tilde{t}$ exists.
\end{proof}

\section{Implementation and Benchmarking}

\subsection{Implementation}
We implemented simulation and training of (n)UDEs using established software packages. Simulation, objective function evaluation and gradient calculation was implemented in the Advanced Multi-language Interface for CVODES and IDAS (AMICI) \citep{frohlichAMICIHighperformanceSensitivity2021}. To ensure scalability, adjoint sensitivity analysis was employed \citep{frohlichScalableParameterEstimation2017}. Parameter estimation problems were specified using the Parameter Estimation Table (PEtab) format \citep{schmiesterPEtabInteroperableSpecification2021}. We work with two pre-existing biologically-inspired models (termed Lotka-Volterra and Boehm), with details provided in the following sections.

As suggested in the literature \citep{hassBenchmarkProblemsDynamic2019}, we $\log_{10}$-transform the mechanistic parameters $\THETAM$ for estimation. The ANN parameters $\THETAU$ are not transformed. $\THETAU$ were generally initialised to be small values and estimated $\in [-10, 10]^{n_\THETAU}$. We used multi-start (1000 starts), gradient-based optimisation with the Fides optimiser \citep{frohlichFidesReliableTrustRegion2022} via the Python Parameter Estimation Toolbox (pyPESTO) \citep{schaltePyPESTOModularScalable2023a}. We initialised the starts by drawing 1000 samples of $\THETAM$ and $\THETAU$, and we reused these 1000 sets of vectors across all comparable experiments, i.e., when using the same model (Lotka-Volterra or Boehm) and ANN architecture.

When assessing whether a solution is negative, we define ``negative'' to be when any state variable in the solution drops below a small negative value, which was chosen to filter for numerical noise. This was -1e-7 for the Lotka-Volterra model and -1e-13 for the Boehm model, which are within one order of magnitude of the absolute simulation tolerances used with each problem.

\subsection{Lotka-Volterra model}
\label{section:lv_setup}
For the demonstration of UDEs, nUDES and the different regularisations, we considered a Lotka-Volterra system describing predator-prey population dynamics. The abundance of prey and predator are denoted by $x_1$ and $x_2$, respectively. The standard Lotka-Volterra ODE system is
\begin{align}
\label{equation:lv_odes}
\begin{rcases}
    \displaystyle\frac{dx_1}{dt} &= \alpha x_1 - \beta x_1 x_2, \\
    \displaystyle\frac{dx_2}{dt} &= \delta x_1 x_2 - \gamma x_2,
\end{rcases}
\end{align}
with mechanistic parameters $\THETAM = (\alpha, \beta, \gamma, \delta)$.

As a test case for modelling unknown mechanisms, we consider that the interaction terms of the system are unknown and replace them with an ANN $\U=(U_1, U_2)$. This yields the UDE system ($N_1 = N_2 = 1$) and nUDE system
\begin{align}
\begin{rcases}
    \displaystyle\frac{dx_1}{dt} &= \alpha x_1 + N_1(x_1) U_1(\UINPUT, \THETAU), \\
    \displaystyle\frac{dx_2}{dt} &= N_2(x_2) U_2(\UINPUT, \THETAU) - \gamma x_2.
\end{rcases}
\label{equation:lv_nudes}
\end{align}

We chose a simple 2/2 ANN for $\U$ with $\tanh$ activation functions in the first layer, and the identity in the output layer. The input $\UINPUT:=\X$ is the state vector.

We initialised each entry of $\THETAM$ ($\alpha$ and $\gamma$) randomly in $[10^{-3}, 10^{3}]$ (uniform distribution), and constrained them by the same bounds during estimation. 
Although we saw a benefit when initialising mechanistic parameters to be small (see Supplemental Material, Section 1), we did not do that here, to avoid bias. Without small start points, approximately 50\% of all start points could not be simulated and optimised, for example due to exponential blow-up.

Synthetic data for training and validation were generated by simulating the system in \eqref{equation:lv_odes} with 100 time units with $\alpha=1.3$, $\beta=0.9$, $\gamma=0.8$, and $\delta=1.8$, and $\X(t=0)=(0.44249296, 4.6280594)$. The first 20 time units of simulated data were used for training, and the next 80 time units for validation. The training data had 15\% multiplicative noise ($\mathcal{N}(\mathbf{0}, 0.15\X)$) added to it, to represent noise-corrupted data. 

\begin{figure}[t]
\centering
\includegraphics[width=0.95\columnwidth]{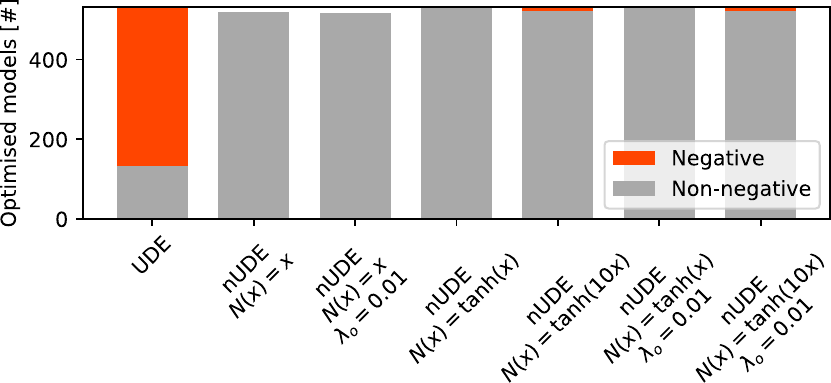}
\caption{Solution characteristics for the Lotka-Volterra model: the number of optimisation runs yielding solutions with negative (orange) or strictly non-negative (gray) values for the predator or prey abundances.}
\label{figure:lv_negatives_predict}
\end{figure}

We simulated each optimisation result to check for non-negativity (Fig. \ref{figure:lv_negatives_predict}). More than half of the UDE fits produced negative populations. All nUDE models had zero negative populations, except with ``$N(x)=\tanh(10x)$''. The ODE solver may have numerical issues near zero due to the larger second derivative of $\tanh(10x)$, than $\tanh(x)$.

\begin{figure}[t]
\centering
\includegraphics[width=0.8\columnwidth]{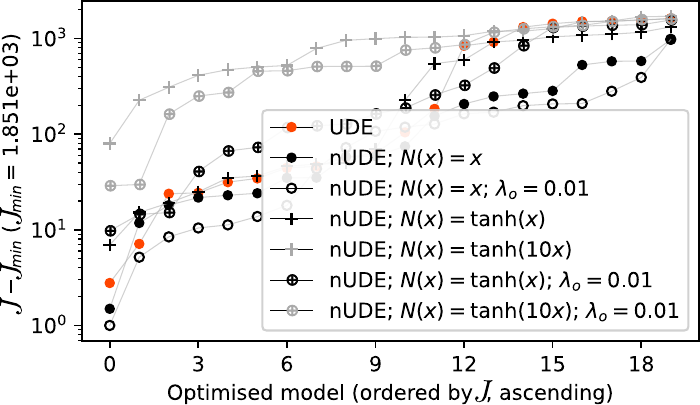}
\caption{Waterfall plot for the Lotka-Volterra model. The objective function value $J$ on the validation data are shown for the 20 best fits on the training data.}
\label{figure:lv_waterfall_predict}
\end{figure}

\begin{figure}[t]
\centering
\includegraphics[width=0.95\columnwidth]{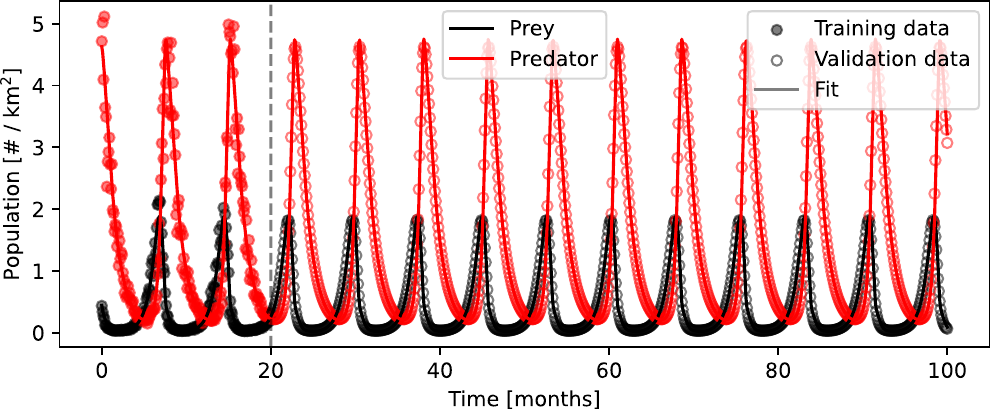}
\caption{Fit and prediction for the Lotka-Volterra UDE. The vertical dashed line indicates the training and validation data split. The fits and predictions from the ``nUDE; $N(x) = x$'' and ``nUDE; $N(x) = x$; $\lambda_o=0.01$'' models are visually indistinguishable to the UDE, and are not shown.}
\label{figure:lv_fits_predict}
\end{figure}

The ``nUDE; $N(x)=x$; $\lambda_o = 0.01$'' case performed best on the training data (not shown), and on the validation data, and the ``nUDE; $N(x) = x$'' was next best on validation data (Fig. \ref{figure:lv_waterfall_predict}). However, the best UDE fit was also very good (Fig. \ref{figure:lv_fits_predict}).

\subsection{Boehm model}

As a second example we consider the Boehm model \citep{boehm2014identification}, which describes the STAT5 dimerisation process. It is fully specified by eight ODEs and nine estimated parameters. The measurements are mapped to the eight state variables through a nonlinear observation function. Implementation and training details are specified in the supplemental materials section 2. To evaluate UDEs and nUDEs, we consider two scenarios for the Boehm model that differ in the effect that the ANN component can have on the overall dynamics: 

\textbf{Scenario 1:} Like in the synthetic Lotka Volterra example, we assume that one mechanism is unknown, here the export and dimer dissociation of \verb|nucpApA|, and remove the term from the ODE. Instead, we introduced a 5/5/5/2 ANN with 82 weight and bias parameters $\THETAU$. This ANN takes only a subset of the state vector as input (specifically, the \verb|nucpApA| species) and modifies the dynamics of two species (\verb|nucpApA| and \verb|STAT5A|).

\textbf{Scenario 2:} A more flexible ANN component is used to emphasise the effect of regularisation. This 5/5/5/5 ANN has the same dimensions in the hidden layers as in \emph{Scenario 1} but takes three state variables as inputs, and modifies the dynamics of five species, which increases the size of $\THETAU$ to 110 free parameters. This scenario represents a greater degree of uncertainty about the missing mechanisms in the model, because the ANN can affect more state variables directly.

\begin{figure}[t]
\centering
\includegraphics[scale=0.45]{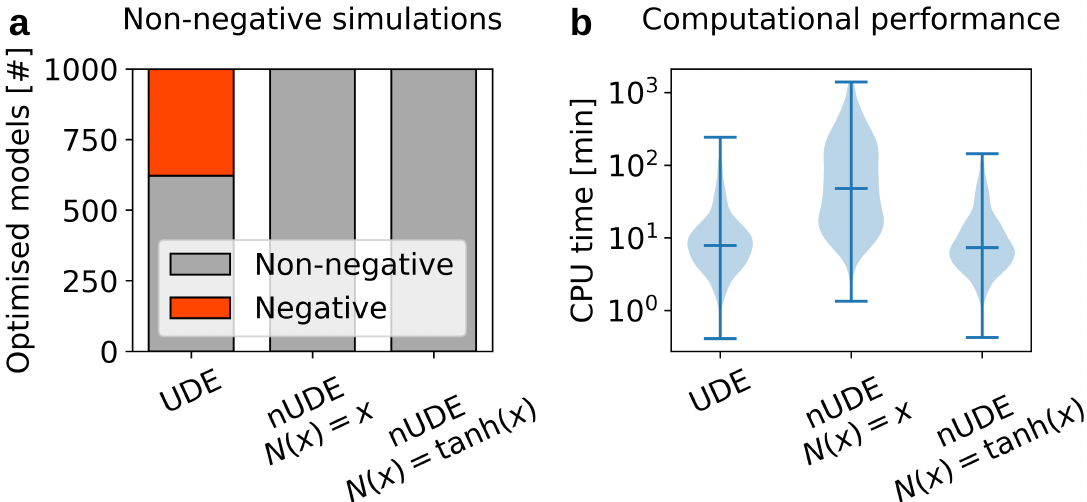}
\caption{Boehm \emph{Scenario 1} comparison between UDE and nUDEs. a) Amount of models that stayed non-negative in their trajectories, and b) distribution of model calibration times per method. Horizontal lines indicate the minimum, medium and maximum.}
\label{figure:boehm_computation}
\end{figure}

\begin{figure}[t]
\centering
\includegraphics[width=0.85\columnwidth]{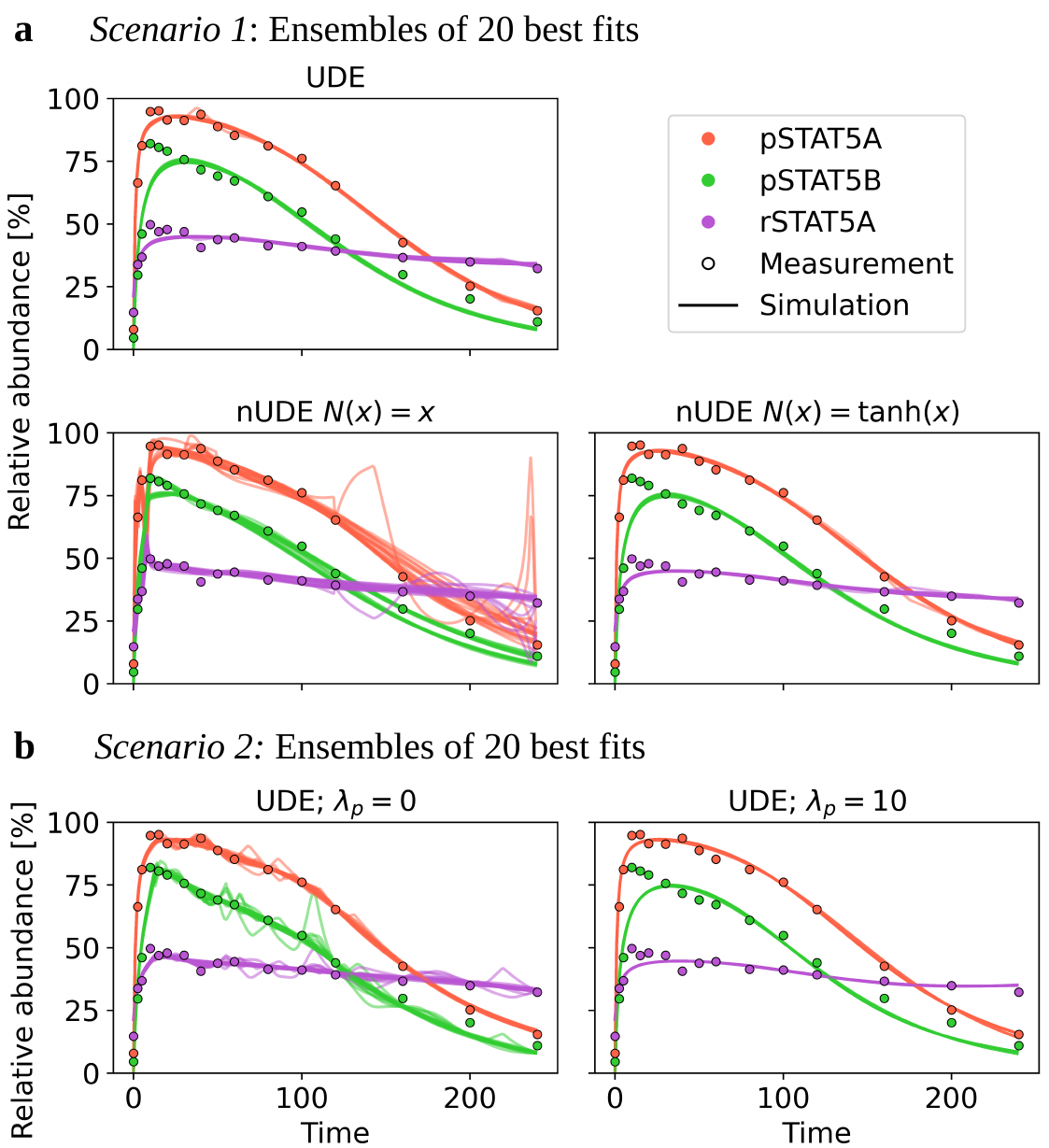}
\caption{Ensembles of 20 best UDE model variants. \textbf{a)} \emph{Scenario 1}: ANN with 1 input/2 outputs. Best fits are shown for UDE, nUDE ($\N(\X)=\X$) and nUDE ($\N(\X)=\tanh(\X)$), no regularisation. 
\textbf{b)} \emph{Scenario 2}: ANN with 3 inputs/5 outputs. Best fits are shown for the unregularised UDE and a parameter-regularised UDE with $\lambda_p=10$.}
\label{fig:boehm_ensembles}
\end{figure}

We first consider \emph{Scenario 1} to assess the non-negativity constraint in a realistic setting.
Of the 1000 starts, 623 optimised UDE fits had non-negative values, while all 1000 nUDE fits were non-negative (Fig. \ref{figure:boehm_computation}a).
However, we found that the overall computation time for optimisation was significantly increased when training the nUDEs with $\N(\X)=\X$ (Fig. \ref{figure:boehm_computation}b). We saw an increased number of optimiser iterations, and simulation time (Supplemental Fig. 3) and furthermore observed predominantly non-smooth trajectories among the best ($\N(\X)=\X$)-nUDE results, indicative of over-fitting (Fig. \ref{fig:boehm_ensembles}a). 
When using $\N(\X)=\tanh(\X)$ however, all 1000 parameterised nUDEs stayed non-negative in their trajectories (Fig. \ref{figure:boehm_computation}a) while the computational cost (Fig. \ref{figure:boehm_computation}b) and the UDE's quality of fit (Fig. \ref{fig:boehm_ensembles}a) were recovered. 
The 2\% of best fits shown in Fig. \ref{fig:boehm_ensembles} agree much better for the ($\N(\X)=\tanh(\X)$)-nUDE than in the ($\N(\X)=\X$)-nUDE, indicating better convergence.

We used the \emph{Scenario 2} UDE variant with a larger ANN component to assess the effect of regularisation. The ANN flexibility has a considerable effect on UDE convergence and over-fitting, as shown by the difference in trajectories between the unregularised \emph{Scenario 1} and \emph{Scenario 2} UDEs (Fig. \ref{fig:boehm_ensembles}a and \ref{fig:boehm_ensembles}b).
With increasing ANN complexity the number of UDEs with negative values increased from 37.7\% in \emph{Scenario 1} to 87.7\% in \emph{Scenario 2}.

There is a substantial difference in the quality of fits between the unregularised and regularised UDEs, as apparent from the best 2\% of fits (Fig. \ref{fig:boehm_ensembles}b). The unregularised UDEs tend to over-fit the training data, characterised by a tight fit to the measurements and a high variability in their trajectories between measurements, with frequent spikes. The regularised UDEs on the other hand have a high agreement between the 20 best models and produce smooth trajectories, as shown for the parameter regularisation in Fig. \ref{fig:boehm_ensembles}b. We observed similar trends between parameter and output regularisation.

\begin{figure}[t]
\centering
\includegraphics[width=0.85\columnwidth]{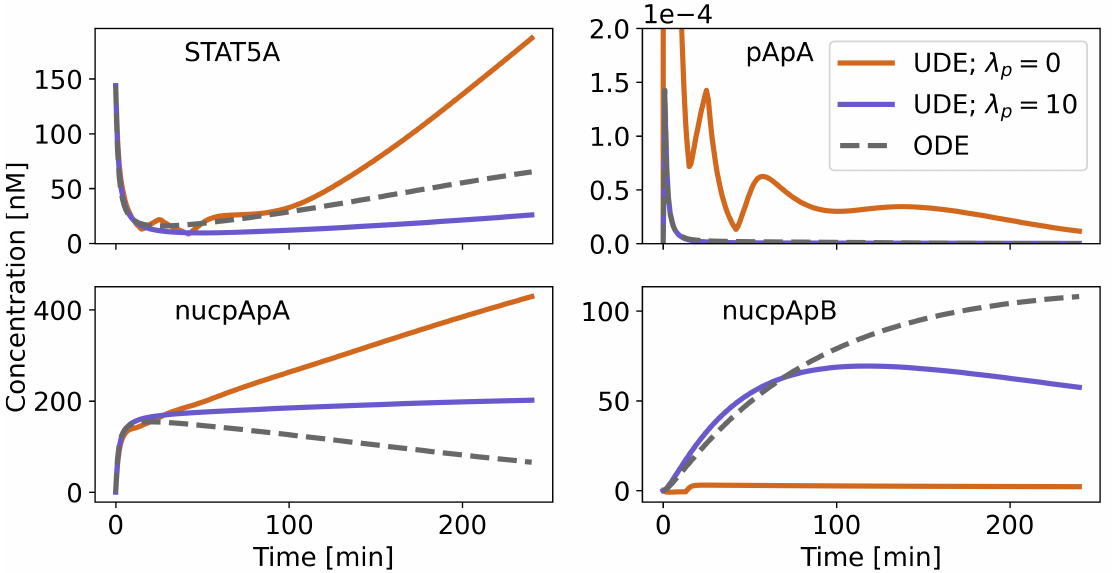}
\caption{Boehm \emph{Scenario 2}: Simulation of best unregularised UDE (orange) and UDE with parameter regularisation ($\lambda_p=10$,  purple). Shown are the simulations for 4/8 state variables.}
\label{fig:boehm_reg_simulations}
\end{figure}

The UDE is expected to give a comparable or closer fit to measurements compared to the fully mechanistic model due to the flexibility of ANNs. However, this is not directly indicative of its generalisation capacity for predictions or inference of non-observed states variables. 
In the real-world Boehm example the reference for the non-observed state variables is not the true solution, which is unknown, but the optimal solution that was obtained from the fully mechanistic ODE model.
Compared to the ODE reference, the normalised mean squared error (NMSE) for the best UDE was significantly improved by parameter and output regularisation (NMSE with $\lambda_p = \lambda_o = 0$: 145.99, NMSE with $\lambda_p = 10$: 12.83, NMSE with $\lambda_o = 0.1$: 12.69). This trend is also discernible in the simulations in which regularisation overcomes the blow-up and strong fluctuations observed in the unregularised UDE (Fig. \ref{fig:boehm_reg_simulations}).

\section{Discussion}

In this manuscript, we presented and evaluated different types of regularisation on the universal components of UDEs describing biological processes. In particular, we introduced (i) an output regularisation to avoid over-fitting, and (ii) a regularisation of the UDE structure to ensure non-negativity.

Our theoretical result guarantees that the solution is non-negative everywhere, up to numerical noise, and this constraint is generally applicable to biological modelling, where entities like molecular concentrations and population sizes are often non-negative. Our experimental results demonstrate that the non-negative constraint works in principle on a synthetic Lotka-Volterra example, and in practice on the real-life Boehm example.

The choice of $\N(\X)$ can itself be mechanistically informed. If there is some prior knowledge that a missing mechanism affecting $x_i$ has the factor $x_i$, then choosing $N_i(\X)=x_i$ may improve the learning problem for $\U$. However, if the missing mechanism does not depend on $x_i$, then $\U$ needs to counter this in addition to learning the missing mechanism. In such cases, we suggest the bounded $N_i(\X)=\tanh(\alpha x_i)$, which does not grow with $x_i$ except near 0, according to $\alpha$. This can have substantial computational benefits (Fig. \ref{figure:boehm_computation}). However, larger choices of $\alpha$ can increase numerical error (Fig. \ref{figure:lv_negatives_predict}), so alternative choices of $\N$ are an important open topic.

We found that $\N$ did ensure non-negativity in computational experiments, but only up to numerical error. Tailored ODE solvers can ensure that user-provided constraints are satisfied by performing additional integration steps as a constraint is approached \citep{eichConvergenceResultsCoordinate1993}. This could be used to remove negativity due to numerical error in nUDEs, but does not resolve the negativity in standard UDEs.

In principle, our regularisation strategies are applicable to a variety of universal approximators, modelling formalisms such as partial differential equations, and choices of $\N$. 
We present some limited benchmarking here. Our results for the Boehm model in \emph{Scenario 2} suggest that, as the amount of prior assumptions on the missing dynamics decreases, the user is forced to choose a more expressive ANN ($\U$), and the importance of regularisation increases. More comprehensive benchmarking is required to uncover best practices when modelling unknown mechanisms.

Hybrid models promise to bridge the gap between the interpretability of mechanistic models, and the predictive capabilities of machine learning models. Context-specific modelling choices can improve the performance of hybrid models substantially. We integrated UDEs into standard workflows for systems biology and showed that biologically-reasonable predictions are possible, without sacrificing computational efficiency.

\begin{ack}
We are grateful to Polina Lakrisenko for fruitful discussions.
Optimisation was performed on the Bonna and Unicorn clusters at the University of Bonn.
\end{ack}

\bibliography{ifacconf}

\appendix

\end{document}